\documentclass[aps]{revtex4}

\usepackage{graphicx}
\usepackage{natbib}

\newcommand{\be}{\begin{equation}}
\newcommand{\ee}{\end{equation}}
\newcommand{\nn}{\mbox{} \nonumber \\ \mbox{} }
\newcommand{\ba}{\begin{eqnarray}}
\newcommand{\ea}{\end{eqnarray}}
\newcommand{\om}{\omega}

\newcommand\eg{\textit{e.g.}}

\newcommand\ie{\textit{i.e.}}

\newcommand{\Bf}{{magnetic field}}
\newcommand{\Bfs}{{magnetic fields}}
\newcommand{\Ef}{{electric  field}}
\newcommand{\Efs}{{electric fields}}
\newcommand{\CR}{{cosmic ray}}
\newcommand{\CRs}{{cosmic rays}}

\newcommand{\E}{{\bf E}}
\newcommand{\B}{{\bf B}}

\renewcommand{\v}{{\bf v}}

\newcommand{\curl}{{\rm curl}}
\renewcommand{\div}{{\rm \,div}}

\newcommand{\jgr}{J. Geoph. Res.}
 \newcommand{\mnras}{MNRAS}
 \newcommand{\apjl}{ApJ Lett.}
 \newcommand{\aap}{AAP}
 \newcommand{\physrep}{Phys. Rep.}
 \newcommand{\apss}{Astroph. Space Sci.}
\begin{document}

\title{Structure of Cosmic Ray-modified Perpendicular Shocks}

\author{Maxim Lyutikov}

\address{
Department of Physics, Purdue University,
 525 Northwestern Avenue,
West Lafayette, IN
47907-2036 }


\begin{abstract}

Kinetic diffusion of cosmic rays ahead of perpendicular shocks induces charge non-neutrality, which is mostly, yet not completely,  screened by the bulk plasma via   polarization drift  current. Hydrodynamic shear instabilities as well as modified  Buneman instability of the polarization  current  generate the turbulence necessary for a   Fermi-type acceleration. Thus, similar to the case of parallel shocks,  in  perpendicular shocks the diffusing cosmic rays generate unstable plasma currents that in turn excite  turbulence. This allows a  self-consistent evolution of a shock-cosmic rays system. In the kinetic regime of the  modified  Buneman instability, electrons may be heated in the cosmic ray precursor.
 \end{abstract}

\maketitle

\section{Introduction}

Acceleration of \CRs\ is one of the main problems of high energy astrophysics. Shock acceleration is the leading model \cite{BlandfordEichler}. 
Particle acceleration at  quasi-parallel shocks  (when the \Bf\ in the upstream medium is nearly   aligned  with  the shock normal) and  quasi-perpendicular shocks
 (when the \Bf\ in the upstream medium is nearly orthogonal  to the shock normal) proceeds substantially differently. 
 Most astrophysical shocks are quasi-perpendicular, yet  theoretically   acceleration at this type of  shocks is less understood than in the case of quasi-parallel  shocks.
  It is recognized that the feedback of accelerated \CRs\  may considerably modify the parallel shock structure  \cite{1982A&A...111..317A}. 

For parallel shocks, the key issues is the generation of magnetic turbulence in the upstream plasma,   required to overcome the particle escape ahead of the shock. A powerful MHD instability is driven by \CRs\ themselves (Bell's instability \citep{Bell04}), providing   a self-consistent description of the \CR\ acceleration, in a sense that that the turbulence and accelerated particle  may reach a steady state. (The transitional period of reaching that state, \ie\ starting  acceleration  from first principles and particle injection, is not addressed by these models). 

In case of perpendicular  (superluminal) shocks, the main problem is advection of particles downstream away from the shock: in such
shocks particles cannot catch up with the shock by streaming along the field lines. Here diffusion in the downstream region is assumed to play a major role. Mostly discussed is a particle diffusion due to field line wandering \cite{1999ApJ...520..204G,Matthaeus}. 

Here we investigate a  model similar in spirit to Bell's \citep{Bell04} approach, but  in  applications to  perpendicular shocks:   assuming that a shock accelerates particles that  diffuse kinetically ({\it not} by field line wandering)  far  upstream,  we investigate the consequence of this assumption.
The key difference between  \CR-modified parallel and   perpendicular shocks is that the \CR\ precursor in perpendicular shocks is non-neutral.
A shock picks up and accelerate some fraction of ISM ions; thus some ions, which were behind the shock if diffusion were absent, are  transported ahead of the shock.
This creates excess of positive charges upstream and a deficit downstream. We stress that we investigate a possibility that ions diffuse many Larmor radii upstream; this is different from the problem of electron heating and acceleration by reflected ions, which occurs on one Larmor radius of reflected ions \citep{1984ZhETF..86.1655G,1988Ap&SS.144..535P},.

In case of parallel shocks such excess of positive charges upstream  is balanced by the electrons drawn along the \Bf\ into upstream plasma, driving Bell's  \citep{Bell04} instability, that excites plasma turbulence and provides scattering centers necessary for Fermi-I acceleration. In contrast,
in  case of  perpendicular shocks, the charge disbalance cannot be easily  cancelled, since kinetic diffusion of  electrons  across \Bf\ is much weaker than that of high energy \CRs.
In principle,  the charge density  induced by \CRs\
can be canceled by electrons coming ''from the sides". But  in the frame  of the incoming plasma, the charge density builds on  very short time scale  of the order of diffusion length  $L$ over shock speed $v_s$, 
so that for sufficiently large transverse dimensions of a shock $l_\perp> v_{T,e} L/v_s$, electrons transported along the \Bf\ at electron thermal velocity $v_{T,e}$ does not have sufficient time to cancel the ion charge.
Thus, even non-relativistic shocks are expected to develop charge density upstream, which, as we show below, play an important role in determining the structure of the \CR\  precursor and possibly leads to turbulence generation.

 Cosmic rays  diffusing ahead of the shock offset the charge balance in the incoming plasma, which becomes non-neutral, with \Ef\ directed along the shock normal. The incoming upstream plasma will {\it partially } compensate this charge density by a combination of electric and polarization drifts. First, the  electric field created by \CRs\ will produce electric drift along the shock surface. All plasma components will drift with the same velocity, so that electron and ion currents cancel each other, while the \CR\ current contributes to enhancement of the \Bf\  in the frame of the shock; \Bf\ in the fluid frame remains the same, see \S \ref{zerom}. Since the electric field increases towards the shock, while  the rest frame \Bf\ remains  nearly constant, the electric drift velocity increases. This bulk acceleration drives  an inertial polarization current along the shock normal, whereby ions slightly accelerate toward the shock, their density decreases, partially compensating the charge density induced by \CRs.
 Thus, 
the charge disbalance  created by \CRs\ is {\it partially} compensated by polarization current in the upstream plasma. 

There is a number of instabilities that can operate in the \CR-modified  upstream plasma. First, there are fluid instabilities due to shear in the upstream plasma. If the flow is inviscid, 
the   Rayleigh-Taylor instability due to variable electric field drift may be excited if there is an  inflection point in the electric drift velocity (Rayleigh criterium). Secondly, small viscosity may drive viscose instabilities, if the Reynolds number is high enough. Finally, current-driven instabilities, in particular of the modified Buneman type, may generate plasma turbulence. (We consider the latter in more details below.) In  all these cases, the resulting instability has wave vector preferentially perpendicular to the initial \Bf, generating the field line wandering required for  acceleration of \CRs\ in the first place. Thus, similar to parallel shocks, assumption of turbulence and \CR\ acceleration leads to turbulence generation by \CRs\ themselves. This indicates that a steady state of \CR-modified shocks is possible.  

\section{Principal issues}

\subsection{Distribution of  \CRs\ ahead of the shock}

Let us neglect, as a first approximation, the  dynamical influence of the \CRs\  on plasma bulk motion. In the frame  of the shock,  plasma with density $n_{0,ISM}$ is moving with constant velocity $v_s$ in the positive $z$ direction. Let the shock be  located at $z=0$, see Fig. \ref{perp-picture}. Consider  \CRs\ in the fluid approximation, described by a local density $n_{cr}$ (this implies that we consider scales much larger than Larmor radii of \CRs). Cosmic rays are advected in the positive $z$ direction with velocity $v_s$ and diffuse with a given diffusion coefficeint $\kappa$. Balancing advection and diffusion,
\be
v_s \partial _z n_{cr}= \kappa\partial _z^2  n_{cr}
\label{nacre}
\ee
we find
\be
 n_{cr} = n_{0,cr} e^{z/L}, \, z< 0
 \label{ncr}
 \ee
 where   $L= \kappa /v_s$ and we assumed that the shock accelerates \CR\ protons with typical density $n_{0,cr}$. A total excess  positive charge of \CRs\ per unit area ahead of the shock, $\sigma = n_{0,cr} L$ is compensated by the similar lack of positive charges behind the shock.
   As a first approximation we neglect changes in the diffusion coefficeint $\kappa$ due changing upstream \Bf,  so that $L$ is a given parameter of the problem.
    As an estimate (an upper limit, in fact) we can use Bohm approximation for the diffusion coefficient,  $\kappa = \gamma c^2/\om_{B,i}$, where $\om_{B,p} = e B/( m_p c)$ is proton cyclotron frequency and $\gamma$ is a typical Lorentz factor of \CRs.  In this case the diffusion length  
    \be
    L=    \gamma c^2/(\om_{B,i} v_s)
    \label{L}
    \ee
    is  larger than ion  Larmor radius in the shocked plasma  $r_L  \approx v_s /\om_{B,i}$ by a factor $\gamma (c/v_s)^2  \gg 1$. Also, diffusion length is much larger than \CR\  Larmor radius by a factor $c/v_s \gg 1$.


Previously,  a number of authors discussed back reaction of ions reflected from the shock into upstream medium \citep[\eg][]{1983ZhETF..85.1232V,1997MNRAS.291..241M}. Reflected ions propagated back into the upstream plasma a distance of the order of one  Larmor radius  $r_L  \approx v_s /\om_{B,i}$ \citep{1996JGR...101.4871G}. This may be important for dynamics  of electrons (\eg, pre-acceleration), having small Larmor radii \cite{1984ZhETF..86.1655G}, but the reflected ions do not affect plasma motion on scales larger than Larmor radius of plasma ions. In contrast, we assume that \CRs\ can kinetically diffuse on much large scales $L \gg r_L$ and investigate the consequences of this assumption. 


\subsection{Heuristic  derivation of upstream dynamics}

\begin{figure}[h!]
\includegraphics[width=\linewidth]{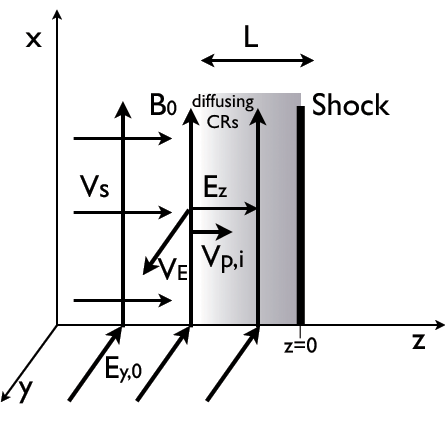}
\caption{Plasma flows in he frame of the shock.
Far upstream, the incoming plasma moves with velocity $v_s$, \Bf\ is along $x$ direction and inductive \Ef\ along $-y$ direction.
At the shock, \CRs\ are accelerated and diffuse ahead of the shock a typical distance $L$, creating an electric field along the shock normal. Electric drift of  \CRs\ in this induced \Ef\   and the  initial \Bf\  produces $z$-dependent electric drift $v_E$  in $y$ direction. Acceleration of plasma in $y$ direction in turn produces polarization drift of ions in $z$ direction.}
\label{perp-picture}
\end{figure}

As a result of \CR\  diffusion ahead of the shock, an  \Ef\ is  created  in the $z$ direction. This \Ef\ will induce a drift velocity along the shock plane, which in turn will induce polarization drift that will partially cancel the initial \Ef. Let the resulting total charge density be $n_{tot}$ and corresponding \Ef\ given by 
$\div E_z = 4\pi e n_{tot}$. Let us neglect variations  of \Bf\ in front of the shock. (A more detailed derivation is given in \S \ref{ddr}.)
The resulting  \Ef\ and the initial \Bf\ induce  a drift velocity in the positive $y$ direction  $v_d = (E_z/B_0) c {\bf e}_y $, see Fig.  \ref{perp-picture}. 
In addition, plasma components will experience polarization drift, which is proportional to mass and thus is mostly important for ions:
\be
v_p = {c\over B  \om_{B,i}} v_s \partial_z  E= {c\over B  \om_{B,i}}  4\pi e n_{tot}= v_s {n_{tot} \over n_{0,ISM}} { \om_{p,i}^2 \over \om_{B,i}^2}
\ee

Continuity of ion flow requires
\be
n_i  (v_s + v_p) = n_{0,ISM} vs, 
\ee
thus creating an excess ion charge
\be
\delta n_i = - {v_p\over v_s} n_{0,ISM}= -  n_{tot}  { \om_{p,i}^2 \over \om_{B,i}^2}
\ee

At any point the total charge is
\be 
n_{tot} = n_{cr}- \delta n_i
= { n_{cr} \over 1+ { \om_{p,i}^2 \over \om_{B,i}^2}} \approx n_{cr} {  \om_{B,i}^2 \over \om_{p,i}^2} \ll n_{cr}
\label{ntot}
\ee
Thus, the polarization current cancels most of the charge induced by \CRs.

Using Eq. (\ref{ntot}) we find the velocity of the polarization drift
\be
v_p = v_s \xi_{cr} { \om_{p,i}^2 \over \om_{p,i}^2+ \om_{B,i}^2} \approx v_s \xi_{cr}
\label{vp}
\ee
Thus, if a fraction of accelerated \CRs\ $ \xi_{cr}$  is small, the polarization drift is much smaller than the shock velocity. 

In case of exponential distribution of \CRs, the resulting \Ef\ is 
\be
E_z \approx  4 \pi  e L { \om_{B,i}^2 \over \om_{p,i}^2} n_{cr} (z)
\ee
and the resulting electric drift is
\be
v_E    \approx  \xi_{cr} (z) L \om_{B,i}  
\ee
If the diffusion length is given by Eq. (\ref{L}),
\be
v_E \approx   \xi_{cr}(z) \gamma {c^2 \over v_s}
\ee

We should also verify that the polarization drift is not affecting strongly the \CR\ distribution.
For the density (\ref{ntot}) the polarization drift of \CRs\ is 
\be
v_{p,cr}  \approx  \xi_{cr}  (z) \gamma v_s
\ee
It can be neglected if $\xi_{cr}  \leq  1/\gamma$. 

The above relations give simple estimates of the effects induced by \CR\ diffusion ahead of the shock. The more detailed derivation is given in the next Section.

\section{Drift motions in non-neutral plasma}

\subsection{General relations}
Let us consider stationary plasma motion in the frame of the shock. Let ion, electron velocities, magnetic and \Efs\ be given by 
\ba &&
\v_i =\{0,v_{i,y}, v_{i,z}\}
\nn &&
\v_e = \{0,v_{e,y}, v_{e,z}\}
\nn&&
\E =\{ 0, -E_{y,0}, E_z\}
\nn&&
\B=\{B_x,0,0\}
\ea
($E_{y,0}$ is constant because $\curl \E=0$).
Assuming massless electrons and massive ions, equations of motion are
\ba &&
\E + {\v_e \over c} \times \B=0
\nn &&
(\v_i \nabla)\v_i = {e \over m_i} \left(\E + {\v_i \over c} \times \B \right) 
\label{1}
\ea
Maxwell's equations and continuity require
\ba &&
\curl \B=  {4\pi e \over c} \left(  n_i \v_i + n_{cr}\left( { E_z \over B_x} c \right)  {\bf e} _y- n_e \v_e \right)
\nn &&
\div \E = 4\pi e ( n_i+n_{cr} - n_e)
\nn &&
n_i v_{i,z} = n_{0,ISM} v_s=n_e v_{e,z}
\ea
Cosmic rays experience just the electric drift along $y$ direction due to combined \Ef\ of \CRs\ and plasma particles. 

Eliminating electron's velocities from Eq. (\ref{1})
\ba &&
v_{e,z} = { E_{y,0} \over B_x} c
\nn &&
v_{e,y}= { E_z \over B_x} c
\ea
we get four equations for  $v_{i,y}, v_{i,z}, B_x, E_z$:
\ba &&
  v_{i,z} v_{i,y} ' = {e \over m_i} \left( - E_{y,0}+ {v_{i,z} \over c} B_x \right)
  \nn &&
  v_{i,z} v_{i,z} '= {e \over m_i} \left(E_z - {v_{i,y} \over c} B_x \right)
  \nn &&
  E_z ' = 4 \pi e \left( n_{0,ISM} v_s \left({1 \over v_{i,z} }  - {B_x \over c  E_{y,0} } \right) + n_{cr} \right) 
  \nn &&
  B_x ' = {4\pi e \over c} \left(  n_{0,ISM} v_s \left({v_{i,y} \over  v_{i,z}}  -{E_z \over E_{y,0}} \right)  + n_{cr} {E_z \over B_x} c \right)
  \label{main}
\ea

\subsection{Zero inertia limit}
\label{zerom}

In this limit we assume that $m_i \rightarrow 0$. This limit is typically not applicable to astrophysical plasmas, since it implies that $\om_{p,i} /\om_{B,i} \rightarrow 0$, while, in fact, an opposite limit, $ \om_{p,i} /\om_{B,i} \gg 1$ is relevant.  Still, the limit of massless protons  allows an exact treatment and highlights a number of important  details. A somewhat different approach to this problem in given in Appendix \ref{N}.

In the massless limit, $m_ i \rightarrow 0$, we find
\ba &&
v_{i,y} = {E_z \over E_{y,0}} v_{i,z}
\nn && 
B_x = { c E_{y,0} \over  v_{i,z}}
\ea
which gives
\ba &&
B_x ^\prime = { 4 \pi e  n_{cr} E_z \over B_x}
\nn &&
E_z^\prime = { 4 \pi e n_{cr}}
\ea

For a given distribution of \CRs, of the form $n_{cr} = n_{0,cr} e^{z/L}$,
we find
\ba &&
v_{i,z} = { 1  \over \sqrt{ 1+ (L/r_B(z))^2} } v_s
\nn  &&
v_{i,y} =  { L/r_B(z) \over \sqrt{ 1+ (L/r_B(z))^2} } c
\nn &&
B_x = B_0  \sqrt{ 1+ (L/r_B(z))^2}  
\nn &&
n_i = n_{0,ISM}  \sqrt{ 1+ (L/r_B(z))^2 } 
E_z = {L\over r_B(z)} B_0
\label{mi0}
\ea
where  
\be
r_B = { B \over 4 \pi  n_{0,cr} e}
\label{rB}
\ee
is the  so-called magnetic Debye radius associated with \CRs. 
(The statement in  \cite{Gordeev06} that  in non-neutral plasmas the  \Bf\ is screened on the magnetic Debye radius $ r_B$ is incorrect). 
Thus, if $L \sim r_B(0)$, the electromagnetic drift can become relativistic.

Note that \Bf\ in the  plasma rest frame $B'=B_z/\Gamma$, where $\Gamma =1/\sqrt{ 1-(v_y^2+v_z^2)/c^2}$  remains constant, $B' = B_0 \sqrt{ 1 - (v_s/c)^2}$ (it would have been more correct to use $B'$ instead of $B_0$ as a parameter of the problem). 

For a typical ISM plasma 
\be
r_B \approx 10^3 {\rm cm}  {1 \over \xi_{cr}} B_{-5} n_{0, ISM}^{-1} 
\label{rB1}
\ee
is very  small even for small \CR\ fraction $ \xi_{cr} \ll 1$. This highlights the fact that even a small charge disbalance  can lead to substantial changes in the  upstream region.

\subsection{Drift equation with polarization current}
\label{ddr}

In the previous section we showed that charge disbalance ahead of the shock can lead to substantial velocities, and thus requires taking  inertial contributions into account.
In Eq. (\ref{main}), eliminating $y$ component of ion velocity
\be
v_{i,y} = \left(  {E_z \over B_x} -  {m_i v_{i,z} v_{i,z}^\prime \over e B_x}\right)  c
\ee
we find
\ba &&
v_{i,z}- v_s = \left( b_x \partial_ z E_z-\partial_ z  b_x  E_z\right)   { m_i c^2 v_{i,z} \over e B_0^2 b_x^2} -
\left(  (\partial_z v_{i,z})^2 + v_{i,z} \partial^2 _z v_{i,z} - v_{i,z} \partial_ z v_{i,z} \partial_ z  b_x \right) {m_i ^2 c^2 v_{i,z} \over e^2 B_0^2 b_x^2}
\nn &&
\partial_ z  b_x = {4 \pi \over B_0^2 b_x} \left( E_z \left(  n_0 ( {v_s \over v_{i,z}}-b_x) + n_{cr} \right) - {m_i n_0 v_s \partial_ z v_{i,z} } \right) 
\nn &&
\partial_ z E_z= 
{4 \pi e}  \left(  n_0 ( {v_s \over v_{i,z}}-b_x) + n_{cr} \right)
\ea
where we introduced $b_x=B_x/B_0$.

The drift approximation corresponds to expansion in  $1/e$, one over electric charge  \cite{Kulsrud}. In a non-neutral plasma it is  important to keep two leading  expansion orders:
\ba &&
v_{i,z}- v_s = \left( b_x \partial_ z E_z-\partial_ z  b_x  E_z\right)   { m_i c^2 v_{i,z} \over e B_0^2 b_x^2} 
\nn &&
\partial_ z  b_x = {4 \pi \over B_0^2 b_x} \left( E_z \left(  n_0 ( {v_s \over v_{i,z}}-b_x) + n_{cr} \right) - {m_i n_0 v_s \partial_ z v_{i,z} } \right) 
\nn &&
\partial_ z E_z= 
{4 \pi e}  \left(  n_0 ( {v_s \over v_{i,z}}-b_x) + n_{cr} \right)
\ea
To proceed further, we assume that the polarization drift is weak, and expand in small quantities $v_{i,z}- v_s$, $b_x-1$, $n_{cr}$ and $E_z$.
\ba &&
v_{i,z}- v_s = - v_s (b_x-1)+ { m_i c^2 v_s \over B_0^2 e} \partial_ z E_z
\nn &&
\partial_ z b_x=- { 4\pi m_i n_{0,ISM} \over B_0^2} \partial_ z v_{i,z}
\nn &&
\partial_ z E_z= 4\pi  e \left(  n_0( { v_{i,z}- v_s \over v_s} -b_x) + n_{cr}  \right)
\ea

Which immediately gives
\be
{v_{i,z}- v_s \over v_s}= -  M_A^{-2} (b_x-1)
\ee
where $M_A = v_s/v_A$. The solutions for the electromagnetic fields and components' velocities are
\ba &&
b_x= 1+ { \xi_{cr}(z) \over (1+\beta_A^2) ( 1-1/M_A^2)} \approx 1+ \xi_{cr}(z)
\nn &&
E_z=  { L \om_{B,i} \over c} { \xi_{cr}(z)  \over 1+\beta_A^2 }B_0 \approx  { L \om_{B,i} \over c} { \xi_{cr}(z)  }B_0
\nn &&
v_{e,y}=  L \om_{B,i} { \xi_{cr}(z)  \over 1+\beta_A^2 } \approx  L \om_{B,i}  \xi_{cr}(z)
\nn &&
v_{e,z}= v_s \left(1 -  { \xi_{cr}(z) \over (1+\beta_A^2) ( 1-1/M_A^2)}  \right)  \approx v_s (1 -  \xi_{cr}(z))
\nn &&
v_{i,y}=  {\om_{B,i} c^2 \over L \om_{p,i}^2}  { 1+ (L \om_{p,i} /c)^2 ( 1-1/M_A^2)\over (1+\beta_A^2) ( 1-1/M_A^2)} \approx L \om_{B,i}  \left( 1+ {c^2 \over \om_{p,i}^2 L^2} \right) \xi_{cr}(z)  
\nn &&
n_e = n_{0,ISM} \left( 1+  {\xi_{cr}(z) \over (1-1/M_A^2)(1+\beta_A^2)}\right)
\nn &&
n_i = n_{0,ISM} \left( 1+  {\xi_{cr}(z) \over (M_A^2+1)(1+\beta_A^2)}\right)
\nn &&
n_i-n_e= - {n_{cr} \over 1+\beta_A^2}
\nn &&
n_{tot} = {\beta_A^2 \over 1+\beta_A^2 } n_{cr}   \approx  \beta_A^2  n_{cr} \ll n_{cr}
\label{mmmain}
\ea
where $\beta_ A = (\om_{B,i}/\om_{p,i}) $.
(Relations (\ref{mmmain}) assume a specific  \CR\ density distribution, $\xi_{cr}(z) = \xi_{cr,0} e^{z/L}$.)
Equations (\ref{mmmain})  solve a problem of modification of the upstream plasma in  perpendicular shocks by cosmic rays  diffusing ahead of the shock. 

The electric drift velocity is typically much smaller than the shock velocity.
If we introduce parameter $\eta$ as cosmic ray acceleration efficiency,
\be
\eta= {  \gamma n_{cr} m_i c^2 \over m_i n_{0,ISM} v_s^2 /2},
\ee
then  with partial screening of the \CR-induced \Ef\ by the incoming plasma, the transverse velocity of the incoming plasma, 
\be
v_E \sim \xi_{cr} c^2 /v_s \approx v_s {  \eta  \over 2} v_s 
\ee
  is typically smaller than the shock velocity.

\section{Instabilities}

\subsection{Fluid instabilities in upstream  plasma}

There is a number of instabilities that can operate in the upstream plasma. First, there are  fluid instabilities
 that may develop in the sheared flow in the upstream medium: KelvinÐHelmholtz (KH) instability of ideal flows and viscously-driven instabilities of flows with high Reynolds numbers. KH instability of sheared flows requires inflection point in the velocity profile 
(Rayleigh condition). This cannot be achieved in the present model, where diffusion coefficient  was assumed to constant, so that electric drift velocity is proportional to smoothly decreasing density of \CRs,   Eq. (\ref{mmmain}). On the other hand, if diffusion has a more complicated spacial dependence (\eg\ it is higher closer to the shock, where the level of turbulence is higher, one might expect an inflection point. 

Allowing for spacial variation of the diffusion coefficient, the \CR\ number density becomes
 \be
 \partial _ z^2 n = {  v- \partial _ z \kappa \over  \kappa}  \partial _ z n
\ee
 Thus, if the diffusion changes on scale $L \sim v/\kappa$, one may have an inflection point in the flow. (Note that since diffusion is lower far away from a shock, $\partial _ z \kappa > 0$.) For example, if the diffusion coefficient has a dependence $\kappa = \kappa_0 e^{z/L}$, the inflection point occurs at 
 $z= - L \ln (\kappa_0 /(Lv_s))$. 
  Eq. (\ref{mmmain}) then indicates that the electric drift velocity will also have an inflection point, resulting in  KH instability. 
 
\subsection{Electromagnetic instabilities: modified Buneman}

In addition to fluid instabilities,  plasmas with ''real'' drift of particles (as opposed, \eg\ to   effective  Larmor drift in inhomogeneous plasma), is subject to a  number of powerful instabilities, related, generally speaking to Buneman current instability, and in particular to the  so called modified  Buneman instability \citep{1962JNuE....4..111B}. 

In keeping with our approach, we next consider hydrodynamic-type (non-kinetic)  instability driven by a perpendicular ion current. Since the  modified  Buneman instability is of the mixed electrostatic-electromagnetic type, it is necessary to keep  electron inertia. Let us transform to a frame  which is drifting  with velocity $v_E$ along the shock normal and is advected with  shock velocity $v_s$. In this frame the ions are drifting with respect to electrons with the polarization drift  velocity $v_p \sim  \xi_{cr} v_s$ directed along $z$ axis. In considering stability of such plasma, we  neglect small variations of the \Bf\ and associated current along $y$ direction, as well as related variations of  electron and ion densities.

Since the  electrostatic contributions to the dielectric tensor are most important, 
we chose  wave vector of perturbations along $z$ direction, $ {\bf k} = k_z  {\bf e}_z$. Let the fluctuating part of the vector potential (in radiation gauge)  ${\bf A}$ be ${\bf A} = a_y {\bf e}_y + a_z {\bf e}_z$. 
The fluctuating electric and \Bfs\ are $\delta \E =  i (\om/c)  \{ 0, a_y,a_z\}$, $\delta \B = -  i k_z \{ a_y ,0,0\}$. 
Introducing electron and ion displacements ${\bf \xi}_e  =   \{\xi_{e,x}, \xi_{e,y},\xi_{e,z}\}$ and ${\bf \xi}_i  =   \{\xi_{i,x}, \xi_{i,y},\xi_{i,z}\}$, 
the equations of motions for electrons and ions are
\ba &&
\partial _t {\v}_e + ( \v_e \cdot \nabla) \v_e = - { e\over m_e }\left( \E +  {\v}_e \times  \B  \right) 
\nn &&
\partial _t {\v}_i + ( \v_i \cdot \nabla) \v_i = { e\over m_i} \left(\E +  {\v}_i \times  \B   \right)  
\nn &&
\v_i =v_p  {\bf e}_z+ \partial _t{\bf \xi}_i
\nn &&
\v_e = \partial _t{\bf \xi}_e
\ea
Electromagnetic fields are related by 
\ba &&
 \partial _t { \B} = - \nabla \times \E
\nn &&
\partial _t \E = \nabla \times \B - {4 \pi \over c} \left( n_i \v_i -n_e \v_e \right)
\ea
Finally, continuity equations give
\ba &&
\partial _t n_e + \nabla ( n_e {\v}_e) =0
\nn &&
\partial _t n_i + \nabla ( n_i {\v}_i) =0
\ea

We find
\ba && 
\xi_{e,x} = \xi_{i,x}=0
\nn &&
\xi_{e,y} = {e \over m_e c} {1 \over \om^2 -\om_{B,e}^2} \left( i\om a_y + \om_{B,e} a_z \right)
\nn &&
\xi_{e,z}={e \over m_e c} {1 \over \om^2 -\om_{B,e}^2} \left(\om_{B,e} a_y -  i \om a_z \right)
\nn &&
 \xi_{i,y} =   { e \over c m_i \om } {1 \over \tilde{\om } ^2 -\om_{B,i}^2}\left(- i \tilde{\om } ^2 a_y + \om_{B,i}  \om a_z \right)
  \nn &&
 \xi_{i,z} =  - { e \over c m_i \om} {\tilde{\om } \over  \tilde{\om } ^2 -\om_{B,i}^2}  \left( \om_{B,i} a_y + i \om a_z \right)
 \nn &&
 \tilde{\om } = \om - k_z v_p
  \nn &&
 n_{e} =  n_0 \left( 1 - i k_z \xi_{e,z} \right)
\nn &&
n_i=  n_0 \left( 1 - i { k_z  \om  \xi_{i,z} \over \tilde{\om }} \right)
 \ea
which gives the equation for normal modes
  \be
    \left(
  \begin {array} {cc}
   n^2 -1 +S_e +  { \tilde{\om } ^2  \over \om^2}  S_i& iD\\
   - i D & -1 + S_e + S_i
   \end{array} 
  \right)
    \left(
  \begin {array} {c}
  a_y \\ a_z
  \end{array}
   \right)=0
  \label{S}
  \ee
  where
  \ba &&
  S_e = {\om_{p,e}^2 \over \om^2 -\om_{B,e}^2}
  \nn &&
  S_i = {\om_{p,i}^2 \over \tilde{\om } ^2 -\om_{B,e}^2}
  \nn &&
  D=-  { \om_{B,e} \over \om} S_e + { \om_{B,i} \over \om} S_i
  \ea
 The  dispersion relation follows from (\ref{S}):
 \be
 n^2=1-S_e -  { \tilde{\om } ^2  \over \om^2}  S_i - {D^2 \over 1- S_e -S_i}
 \ee
 The eigenmodes satisfy
 \be
 a_y= i { 1- S_e -S_i \over D} a_z
 \label{eigen}
 \ee
 Thus, the eigenmodes are elliptically polarized. Note, that  the wave vector of the perturbations lies in the polarization plane, so that the wave is of a  mixed  electrostatic-electromagnetic type.

For vanishing drift velocity, $v_p=0$,  in the low frequency limit $\om_{B,e} \rightarrow \infty$,  Eq.  (\ref{S}) gives  the extraordinary mode
\be
n^2= 1-S_e -S_i -{ D^2 \over 1 -S_e -S_i} = 1+{ ( \om_{p,i}^2+ \om_{B,i}^2)\om_{p,i}^2  \over \om_{B,i}^2 ( \om_{p,i}^2+ \om_{B,i}^2 - \om^2) } \approx  \om_{p,i}^2/\om_{B,i}^2
\ee

In case of finite drift velocity, keeping terms linear in $m_e/m_i$ and assuming $   \tilde{\om } \ll  \om_{B,i} $, we find
\be
n^2 =1 + \left( {m_e \over m_i}  + {  \tilde{\om } ^2 \over \om^2} \right) {\om_{p,i}^2 \over  \om_{B,i}^2}
\label{Disp}
\ee
which gives
\be
\om \approx k_z  \left( v_p \pm \sqrt{ v_A^2 - v_p^2 ( m_e/m_i +\beta_A ^2) }\right)
\label{omin}
\ee
Equation (\ref{Disp}) has complex roots provided
\be
v_p > {v_A \over \sqrt{ (m_e/m_i) +\beta_A ^2}}
\ee
If $\beta_A  \ll  \sqrt{m_e/m_i}$ (this requires $B < 7 \times 10^{-5} {\rm G} \, n_{0,ISM}^{1/2}$),  the instability occurs for 
\be
v_p > \sqrt{m_i/m_e} v_A
\label{vcrit}
\ee
  Since polarization drifts velocity  is $v_p \sim \xi _{cr}  v_s$, it is required for instability that 
  \be
  \xi _{cr} > \sqrt{m_i \over m_e} {1 \over M_A}
  \label{xi11}
  \ee
  where $M_A =v_s /v_A \gg1 $ is the shock Mach number. 
  
  For velocities larger than (\ref{vcrit}), the instability growth rate is 
  \be
 \Gamma =\sqrt{ m_e \over m_i} \xi _{cr} k_z v_s
 \label{Gamma}
 \ee
 
 The instability has time to grow $\sim L/v_s$;  the requirement $ \Gamma  L/v_s > 1$ limits the wave number of growing modes to 
 \be
2 \sqrt{m_i/m_e} {1 \over M_A \eta} < k_z  \delta_i < 1
 \ee
 where $\delta = c/\om_{p,i}$ is ion skin depth.  The upper limit on $k_z < 1/\delta_i$ follows from neglect of kinetic effects. 
 
 Thus, if  \CR\ acceleration efficiency is
 \be 
 \eta >  2 \sqrt{m_i\over m_e} {1 \over M_A} 
 \label{eta}
 \ee
 instability will have enough time to grow. Condition (\ref{eta}) favors strong shocks with weak \Bf. 
 
 What is the nature of the unstable modes? Eq. (\ref{omin}) indicates that both eigenmodes  become unstable, implying that  both 
  components of the electric fields orthogonal to the initial \Bf, as well as fluctuations of \Bf\ along the initial \Bf, grow exponentially. Eq. (\ref{eigen}) indicates that the phases of 
 the  electric field components $E_z$ and $E_y$ are shifted by ninety degrees, implying that growing modes correspond to vortical perturbations. 
 These kinds of perturbations will shuffle field line along $z$ direction. If we allow for finite perturbations along $x$ and $y$ direction, the instability will induce field line wandering. 
Since   field line wandering can bring both signs of charge, this might be break down our assumption of charge non-neutral plasma. Still, the motion of electrons and ions along the field line might proceed in different regime: resulting magnetic field amplification and corresponding magnetic bottles will reflect electrons more efficiently due to their
 small Larmor radius.
 
 \subsection{Applicability and saturation}

The applicability of our approach requires that (i) the diffusion length $L$ is   larger than the  Larmor radius of \CRs\  $r_{L,cr} =\gamma c/\om_{B,i}$ -- this condition is satisfied for any non-relativistic shock,  since $L \sim (c/v_s) r_{L,cr} \gg r_{L,cr}$; (ii)  $L$ is larger than skin-depth of \CR\ component, $\delta_{cr} = c/\om_{p,cr}$. We find
\be
{ L \over \delta_{cr} } = \sqrt{ \gamma \xi_{cr}} {c\over v_s} {c\over v_A} = \sqrt{\eta \over 2} {c\over v_A}
\ee
This puts  a lower limit on \CR\ efficiency
\be
\eta > 2 (\beta_A)^2= 10^{-10}  {1 \over n_{ISM} } {B_{-6} }^2 
\ee
which can be easily satisfied.

 The growth rate (\ref{Gamma}) was calculated assuming cold ions and electrons. This is justified if  the polarization drift velocity is larger than electrons thermal speed $v_{T,e}$, 
 $v_p= \xi_{cr} v_s > \sqrt{T/m_e}$. This requires 
 \be 
 \xi_{cr} > \sqrt{m_i \over m_e}{1\over M_s}
\label{elec}
 \ee
 where $M_s = v_s /v_{T,i}$ is the conventional Mach number.
  The condition (\ref{elec}) is quite restrictive, so that the modified Buneman instability may also heat electrons. Previously, it was suggested  \cite{2007ApJ...654L..69G,2008ApJ...684..348R} that lower hybrid waves in the CR precursor 
of a perpendicular shock might be a plausible electron heating 
mechanism. To assess the efficiency of electron heating, kinetic-type calculations are required; we postpone them to a future paper. For  drift velocities smaller than ion thermal speed, $\v_p \leq v_{T,i} $  it is expected that the instability will be stabilized.  This requirement places a lower limit on \CR\ faction, $\xi_{cr} >  1/M_A$.

To estimate the saturation level of instability, we assume that a large fraction of the ion polarization drift is converted into fluctuating \Bf. This gives an estimate
\be
\left( { \delta  B \over B_0} \right)^2 \sim M_A^2 \xi_{cr}^3 \geq {1 \over M_A}
\ee

\section{Discussion}

We considered how \CR\ accelerated at perpendicular shocks and diffusing kinetically ahead of the shock modify the upstream flow. 
First, the electric charge density induced by \CR\ diffusion is mostly, but not completely, screened by the polarization drift in the upstream plasma. The remaining charge density induces shear along the shock plane, so that the shock become oblique (in a sense that the flow velocity is not aligned with the shock normal; \Bf\ remains orthogonal to the shock normal).

 For fast shocks with high efficiency of \CR\ acceleration (so that the condition (\ref{xi11}) is satisfied), modified Buneman instability  develops. 
 In addition, a  sheared flow can be unstable to fluid shearing instabilities.  Both fluid and  drift current instabilities would  generate turbulence  and presumably enhanced particle diffusion, vindicating the assumption that perpendicular shocks are efficient \CR\ accelerators.
 The proposed instability can also be important for electron heating in the \CR\ precursor. This requires a kinetic treatment of the instability.
In addition, since the higher energy particles are diffusing faster,   the resulting   population inversion ahead of the shock may drive cyclotron instabilities; we leave consideration of this possibility to a future paper.

In case of parallel shocks, development of Bell's instability leads to generation of fluctuating perpendicular \Bfs, which become dominant over the initial parallel field \cite{2008ApJ...682L...5S}. In that case, the instability discussed in this paper can be regarded as a secondary instability of the \CR-modified parallel shocks.

The effect discussed in this paper is due to non-neutrality of the upstream plasma and thus is easily missed in a conventional MHD treatment. In addition, we assume that kinetic \CR\ diffusion dominates over field line wandering, at least upstream of the shock. If particle were to diffuse mostly by    field line wandering, charge neutrality would  be established by drawing a parallel current. Also, we did not address the issues of particle injections, \ie, how to start the process going. 

I would like to thank Elena Amato, Benjamin Chandran,  Martin Laming and Anatoly Spitkovsky for the most enlightening discussions.

 \bibliographystyle{plain}


\appendix

 \section{Magnetic field in the upstream plasma}
 \label{N}
 
 In this Appendix we re-derive  behavior of \Bf\ ahead of the shock neglecting inertial effects. Let us assume that a  flow carrying \Bf\ $B_0$ approaches a perpendicular shock, and that due to the acceleration of \CR\ particles there is a layer of uncompensated charge in front of the shock. 
  Neglecting diamagnetic currents induced by \CRs\ as well as  pressure of \CRs, the   equation of motion of \CRs\  and electromagnetic fields reduces to
 \be
 \E+{ \v \over c} \times\B =0
 \ee
 here $\E$ and $\v$ are total electric fields and drift  velocities. 
Let  the density  within the  charged layer  be $n_{tot} $  and the thickness of the layer be  $L$. 
 Since the charge-separated current is ${\bf j} = e n_{tot} \v$, and $ \nabla \times \B = (4\pi/c){\bf j}$, we find
 \be
 \E+ {1 \over 4 \pi   e n_{tot} } \nabla \times  \B \times \B=0
 \ee
 Taking divergence of this equation gives
 \be
   n_{tot} + {c^2 \over 16 \pi^2   e^2} \nabla \cdot \left({  \nabla \times  \B \over n _{tot}}  \times \B\right) =0
   \label{2}
 \ee
  The  Eq. (\ref{2}) reminds of the plasma dynamics in the limit of electron MHD \citep{Kingsep}. For  given $n_{tot}$  and  $L$, the modification of \Bf\ depends only on number density of \CRs, Eq. (\ref{2}), and not on their energy.
 
 Eq. (\ref{2}) can be integrated,
 \ba &&
  B_x^2={16  \pi^2   e^2 \over c^2}  { {\cal N}^2 } +   C_1 {\cal N} +C_0
 \nn &&
{\cal {N}} (z) = \int_z^0 n_{tot} (z) dz
\ea
where $C_1$ and $C_0$ are integration constants. 
The quantity ${\cal N} (z) $ is a  surface density of \CRs\ ahead of the shock, integral of density from a given position $z$ up
to the shock front. (Recall that in the zero inertia limit there is no compensating polarization drift of bulk ions.)  In the particular case when the density of \CRs\
is  given by  Eq. (\ref{ncr}), ${\cal N} (z) =  n_{0,cr} L (1-e^{z/L}) $, so that ${\cal N} (z= - \infty)=n_{0,cr} L $. Since the  \Bf\ at $z= - \infty$ is $B_0$, this gives
\be
B_0^2 ={16 \pi^2   e^2  n_{0,cr} L \over c^2}  + C_1 n_{0,cr} L +C_0
\ee
Assuming  assume that $z$-component of the \Ef\ is zero  and \Bf\ is $B_0$ at $z= - \infty$, we find
\ba &&
B_x^2=B_0^2\left( 1 +  { L^2  \over r_B(z)^2}   \right)
\nn &&
E_z = {L\over r_B(z)} B_0  
\ea
Consistent with Eq. (\ref{mi0}).
Again, the apparent  increase of \Bf\ is, in fact,  due to the Lorentz boost along  $y$ direction with electric drift  velocity
\be
{ v_E \over c} = { L /r_B(z) \over  \sqrt{ 1+ L^2 / r_B(z)^2}}
\ee


\end{document}